# Theoretical design of all-carbon networks with intrinsic magnetism

Yan Gao[1], Xiaolong Feng[2], Ben-Chao Gong[1], Chengyong Zhong[3], Shengyuan A. Yang[2,4], Kai Liu[1]†, and Zhong-Yi Lu[1]†

[1]Department of Physics and Beijing Key Laboratory of Opto-electronic Functional Materials & Micro-nano Devices, Renmin University of China, Beijing 100872, China

[2]Research Laboratory for Quantum Materials, Singapore University of Technology and Design, Singapore 487372, Singapore

[3]Institute for Advanced Study, Chengdu University, Chengdu 610106, China

[4]Center for Quantum Transport and Thermal Energy Science, School of Physics and Technology, Nanjing Normal University, Nanjing 210023, China

†Corresponding authors:

kliu@ruc.edu.cn; zlu@ruc.edu.cn

# Abstract

To induce intrinsic magnetism in the nominally nonmagnetic carbon materials containing only *s* and *p* electrons is an intriguing yet challenging task. Here, based on first-principles electronic structure calculations, we propose a universal approach inspired by Ovchinnikov's rule to guide us the design of a series of imaginative magnetic all-carbon structures. The idea is to combine the differently stacked graphene layers via the acetylenic linkages ($-C \equiv C-$) to obtain a class of two-dimensional (2D) and three-dimensional (3D) carbon networks. With first-principles electronic structure calculations, we confirm the effectiveness of this approach via concrete examples of double-layer ALBG-$C_{14}$, triple-layer ALTG-$C_{22}$, and bulk IALG-$C_{30}$. We show that these materials are antiferromagnetic (AFM) semiconductors with intralayer Néel and interlayer AFM couplings. According to the above idea, our work not only provides a promising design scheme for magnetic all-carbon materials, but also can apply to other π-bonding network systems.

## 1. Introduction

Carbon is known for its abundance and its ability to form enormous allotropic structures. Within the carbon family, the magnetic carbon materials are of special fundamental and technological importance. They provide a platform for studying the magnetism from only *s* and *p* electrons and have potential applications in spintronic devices [1-3]. Nevertheless, magnetic carbon allotropes are extremely rare, because intrinsic magnetism tends to be suppressed in most pristine carbon materials (like graphene and diamond) due to the ultra-stable ground state with all electrons paired up. So far, the proposed magnetism in carbon materials largely arises from the structural defects, boundary effects, and other extrinsic origins [4-6]. For example, in two-dimensional (2D) graphene sheets, magnetism can be introduced by special edge geometry [7, 8], defects [9], atomic substitution [10], or partial hydrogenation [11]. Notably, robust ferromagnetism was predicted in half-hydrogenated graphene ($C_2H$) [12], and the magnetic signal has indeed been detected in experiment [13]. However, the strategy relies on the introduction of H adatoms to generate unpaired spins. It is still a challenge to explore approaches that can realize magnetism in pristine carbon allotropes.

It is known that the $2s^22p^2$ orbitals of C atoms in graphene hybridize into three $sp^2$ orbitals and one *p* orbital ($p_z$-orbital) perpendicular to the graphene plane [14]. The $sp^2$-orbitals form strong $\sigma$-bonds as the skeleton of the honeycomb lattice which contains two sublattices labeled as A and B, and the unsaturated $p_z$-orbitals are paired with each other to form a conjugated $\pi$-bonding network covering the A and B sublattices. Here, the number of the unsaturated $p_z$-orbitals on the A sublattice ($N_A^{p_z}$) is equal to that on the B sublattice ($N_B^{p_z}$) in graphene, which guarantees the robust nonmagnetic ground state. It is worth noting that Ovchinnikov [15, 16] proposed that in a conjugated system divided into two groups: $A^+$ and A (that is, each $A^+$ atom is surrounded only by A atoms, and vice versa), it is confirmed that the ground state of the conjugated system should have the total spin $S = \frac{|n_A - n_{A^+}|}{2}$, where $n_A$ and $n_{A^+}$ are the numbers of carbon atoms in groups A and $A^+$, respectively. Inspired by the non-magnetic $p_z$ orbital physics in graphene and Ovchinnikov's proposal [15], then, a natural idea to generate magnetism is to destroy the equality of the unsaturated $p_z$-orbitals of the A and B sublattices on the graphene sheet, thereby inducing the net magnetic moment $\mu_B |N_A^{p_z} - N_B^{p_z}|$, where $N_A^{p_z}$ and $N_B^{p_z}$ represent the numbers of the unsaturated $p_z$ orbitals on the A and B sublattices of the graphene sheet, respectively.

In this work, we develop the above idea into a universal design principle for generating both two-dimensional (2D) and three-dimensional (3D) magnetic carbon materials. We first test the idea in a partially hydrogenated graphene — single-layer $C_6H$ and show that the induced magnetic moment conforms with the above relation. Then, we construct carbon networks by stacking the graphene sheets, and the H's are replaced by the acetylenic linkages ($-C \equiv C-$) which connect the graphene layers. Here, we consider three concrete examples obtained by this strategy: double-layer ALBG-$C_{14}$, triple-layer ALTG-$C_{22}$, and bulk IALG-$C_{30}$. We systematically investigate their stability, electronic and magnetic properties. We show that ALBG-$C_{14}$ and ALTG-$C_{22}$ are 2D antiferromagnetic semiconductors (respective direct bandgaps of 0.32 and 0.31 eV) with intralayer Néel state and interlayer AFM coupling. And IALG-$C_{30}$ is a 3D bulk magnetic carbon allotrope, with properties similar to those of ALBG-$C_{14}$ and ALTG-$C_{22}$.

## 2. Computational methods

To explore the electronic and magnetic properties of carbon networks, the density functional theory (DFT) calculations were carried out with the projector augmented wave (PAW) method [17] as implemented in the VASP package [18]. The generalized gradient approximation (GGA) of the Perdew-Burke-Ernzerhof (PBE) type [19] was adopted for the exchange-correlation functional. The kinetic energy cutoff of the plane-wave basis was set to 550 eV. The slabs with a vacuum space of 20 Å were used to simulate the 2D structures. The $10 \times 10 \times 1$ and $4 \times 6 \times 10$

$k$-point meshes [20] were taken for the sampling of the 2D and 3D Brillouin zones (BZs), respectively. The Grimme's PBE-D2 method [21] was adopted to account for the interlayer van der Waals (vdW) interactions. The energy and force convergence criteria were set to $10^{-6}$ eV and 0.001 eV/Å, respectively. To examine the dynamical stability, the phonon calculations were performed by using the finite displacement method as implemented in the Phonopy package [22], in which the $3 \times 3 \times 1$ and $2 \times 2 \times 2$ supercells are employed for the 2D and 3D structures, respectively. The thermal stability was evaluated with the *ab initio* molecular dynamics (AIMD) simulations in a canonical ensemble [23] with a Nose-Hoover thermostat [24].

## 3. Results and discussion

To test out our idea, we first consider a graphene layer (Fig. 1(a)) with a partial coverage of hydrogen atoms, which is termed as $C_6H$ (Fig. 1(b)). The primitive cell (black dotted line) and the supercell (orange dotted line) of $C_6H$ are shown in Fig. 1(c). We have checked the stability of the material (which will be discussed in a while). Different magnetic configurations are tested, including nonmagnetic (NM), ferromagnetic (FM), AFM Néel (Néel), stripe AFM (Stripe), and zigzag AFM (ZigZag) states, as shown in Fig. 1(j). We find that $C_6H$ always gets optimized to the Néel state, no matter what the initial magnetic configuration is. The energy of the Néel state is 291.3 meV per formula unit (f.u.) lower than that of the NM state, which implies that the intralayer Néel order is very robust against other magnetic orders (see Table 1).

The emergent magnetism in $C_6H$ meets our expectation. Note that $C_6H$ is obtained by hydrogenating one carbon atom from a particular sublattice in a graphene supercell containing six carbon atoms. The adsorption of a H atom will cause a strong $\sigma$-bond formed between the H atom and the C atom beneath it. This means that the $p_z$-orbital (a dangling bond) on the C site is saturated, causing the non-equal number ($m$) of $p_z$-orbitals between the A and B sublattices (here $m = |N_A^{p_z} - N_B^{p_z}| = |3 - 2| = 1$).

The analysis implies that $C_6H$ will possess $1\,\mu_B$ magnetic moment per unit cell. Indeed, we find that the adsorption of H atom leads to a staggered AFM Néel state amounting to exactly $1\,\mu_B$/cell, which confirms the validity of our aforementioned mechanism.

Next, we will get rid of the H atoms to obtain a pure magnetic carbon material. This is achieved by stacking the graphene layers and connecting the layers with acetylenic linkages ($-C \equiv C-$). The essential point is that for each layer, the acetylenic linkages will play the role of the H atoms as in monolayer $C_6H$. Thus, magnetism can be induced in each layer and coupled through the linkages across the layers. To be specific, in this work, we consider three concrete examples: (1) acetylenic-linked bilayer graphene, named as ALBG-$C_{14}$ (Fig. 1(e)), (2) A-B-A-stacked trilayer graphene via acetylenic linkages, named as ALTG-$C_{22}$ (Fig. 1(g)), and (3) a 3D bulk carbon structure termed as IALG-$C_{30}$ (Fig. 1(h)). The unit cells of ALBG-$C_{14}$ and IALG-$C_{30}$ are illustrated in Fig. 1(f) and 1(i), respectively. It's worth noting that in our previous paper [25], with increasing width n of the armchair nanoribbons in the graphene-like plane, a series of nonmagnetic (NM) grapheaynes, which are hybrid *sp-sp$^2$-sp$^3$* carbon allotropes consisting of graphene and acetylenic linkages ($-C \equiv C-$), will be produced, such as grapheayne-1, 2, 3 ⋯ n. Following the same way, based on the three examples here, a series of magnetic 2D and 3D carbon network structures can also be obtained.

For each of the three examples, we investigate its magnetic ground state by comparing different magnetic configurations. We find that they are magnetic, and all prefer the magnetic configuration with the intralayer Néel state and the interlayer AFM coupling (Interlayer-AFM), as illustrated in Fig. 1(j). The detailed comparison is shown in Table 2. The key structural parameters and basic properties of ALBG-$C_{14}$, ALBG-$C_{22}$, and IALG-$C_{30}$ are summarized in Table 3. For comparison, several other typical 2D and 3D carbon allotropes, including T-carbon [26, 27], carboneyane [28], graphyne [29, 30], diamond [31-33], graphite [33], and graphene [34], are also computed and listed. Our calculations indicate that the total energies of ALBG-$C_{14}$

(-8.83 eV/C), ALBG-$C_{22}$ (-8.81 eV/C), and IALG-$C_{30}$ (-8.81 eV/C), although being slightly higher than those of diamond (-9.09 eV/C), graphite (-9.21 eV/C), and graphene (-9.22 eV/C), are lower than those of the experimentally synthesized T-carbon (-7.92 eV/C), the carboneyane (-8.27 eV/C) with $sp$-$sp^2$-$sp^3$ chemical bonds, and the graphyne (-8.58 eV/C) that has been considered to be the most stable in the 2D graphyne family containing the acetylenic linkages ($-C \equiv C-$) [29].

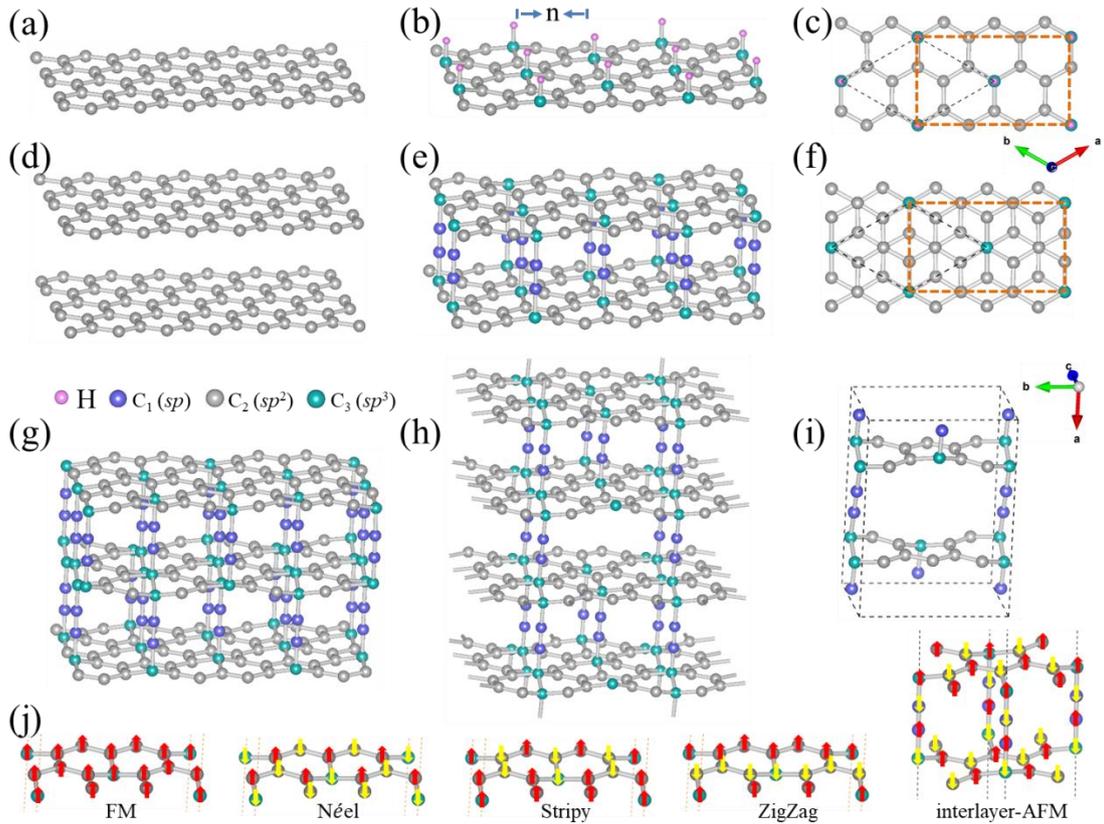

**Fig. 1.** (Color online) Optimized geometric structures and different magnetic configurations. (a) The side view of graphene. (b) The perspective view and (c) top view with the primitive cell (black dotted line) and the supercell (orange dotted line) of $C_6H$. The width n of the armchair nanoribbons in the graphene-like plane is also shown in Fig. 1(b), here the width n equals to 2. (d) The side view of AB-stacked bilayer graphene. (e) The perspective view and (f) top view with the primitive cell (black dotted line) and the supercell (orange dotted line) of ALBG-$C_{14}$. (g) The side view of ALTG-$C_{22}$. (h) The perspective view and (i) the unit cell of IALG-$C_{30}$. (j) Four possible magnetic patterns for the graphene-like plane, namely ferromagnetic (FM), AFM Néel (Néel), stripe AFM (Stripe), and zigzag AFM (ZigZag) states with an interlayer AFM coupling.

To evaluate the stability of these 2D and 3D carbon structures, we have calculated their phonon spectra. As shown in Fig. 2, there is no soft mode in the phonon spectra, indicating that these structures are dynamically stable. The thermal stability is further assessed by the AIMD simulations. We have carried out the simulation at the temperature of 300 K for 10 ps with a time step of 1.0 fs, and confirmed that these materials maintain their integrity against the thermal fluctuations at room temperature. We have also performed the calculations of the elastic constants to assess the mechanical stability of IALG-$C_{30}$. The calculated $C_{11}$, $C_{22}$, $C_{33}$, $C_{44}$, $C_{55}$, $C_{66}$, $C_{12}$, $C_{13}$, $C_{23}$, $C_{15}$, $C_{25}$, $C_{35}$, and $C_{46}$ are 589.1, 559.5, 194.5, 16.5, 16.7, 256.0, 90.5, 37.9, 53.4, 10.7, 12.8, 17.8, and 3.2 GPa, respectively, which satisfy the criteria of mechanical stability for monoclinic phase [35] and confirm that IALG-$C_{30}$ is mechanically stable.

**Table 1**. Nonmagnetic (NM), ferromagnetic (FM), stripe AFM (Stripe), zigzag AFM (ZigZag), AFM Néel (Néel), and interlayer AFM (Interlayer-AFM) for initial states, final states, $\Delta E_{total}$ ($E_{total}$ relative to that of the lowest-energy magnetic order), cohesive energies $E_c$ (eV/C), and $E_c - E_{AFM}$ ($E_c$ relative to that of the lowest-energy magnetic order) of $C_6H$ and ALBG-$C_{14}$, respectively. For ALBG-$C_{14}$, the FM, Stripe, ZigZag, and Néel states represent the corresponding intralayer states with the interlayer FM coupling, and the Interlayer-AFM represents the intralayer Néel state and the interlayer AFM coupling.

| Structures | Initial states | NM | FM | Stripy | Zigzag | Néel | Interlayer-AFM |
|---|---|---|---|---|---|---|---|
| $C_6H$ | Final states | NM | Néel | Néel | Néel | Néel | ~ |
|  | $\Delta E_{total}$ (meV/f.u.) | 291.3 | 0.3 | 0.0 | 0.0 | 0.0 | ~ |
| ALBG-$C_{14}$ | Final states | NM | Néel | NM | Néel | Néel | Interlayer-AFM |
|  | $E_c$ (eV/C) | -8.8095 | -8.8232 | -8.8098 | -8.8232 | -8.8232 | -8.8267 |
|  | $E_c$-$E_{AFM}$ (meV/C) | 17.2 | 3.5 | 16.9 | 3.5 | 3.5 | 0.0 |

**Table 2**. Nonmagnetic (NM), ferromagnetic (FM), and interlayer AFM based on the intralayer Néel couplings (Interlayer-AFM) for initial states, final states, cohesive energies $E_c$ (eV/C), and $E_c - E_{AFM}$ ($E_c$ relative to that of the Interlayer-AFM) of ALTG-$C_{22}$ and IALG-$C_{30}$, respectively. The Néel states represent the corresponding intralayer states with the interlayer FM coupling.

| Structures | Initial states | NM | FM | Interlayer-AFM |
|---|---|---|---|---|
| ALTG-$C_{22}$ | Final states | NM | Néel | Interlayer-AFM |
| | $E_c$ (eV/C) | -8.7901 | -8.8076 | -8.8076 |
| | $E_c$-$E_{AFM}$ (meV/C) | 17.5 | 0.0 | 0.0 |
| IALG-$C_{30}$ | Final states | NM | Neel | Interlayer-AFM |
| | $E_c$ (eV/C) | -8.8010 | -8.8048 | -8.8064 |
| | $E_c$-$E_{AFM}$ (meV/C) | 5.4 | 1.6 | 0.0 |

**Table 3**. Crystal structures, space groups, lattice parameters (Å), angles (°), densities (g/cm$^3$), bond lengths (Å), bulk moduluses (GPa), total energies $E_{tot}$ (eV/C) of ALBG-$C_{14}$, ALTG-$C_{22}$, IALG-$C_{30}$, T-carbon, carboneyane, graphyne, diamond, graphite, and graphene obtained from our work and other works are given for comparison.

| Structures | Space groups | Method | Lattice parameters(Å) | | | Angles (°) | | | Densities (g/cm$^3$) | Bond lengths(Å) | Bulk moduluses (GPa) | $E_{tot}$ (eV/C) |
|---|---|---|---|---|---|---|---|---|---|---|---|---|
| | | | a | b | c | α | β | γ | | | | |
| ALBG-$C_{14}$ | $P\bar{3}1m$ | Our work | 4.31 | 4.31 | ~ | 90.00 | 90.00 | 120.00 | ~ | 1.22-1.51 | ~ | -8.83 |
| ALTG-$C_{22}$ | $C2/m$ | Our work | 4.30 | 7.47 | ~ | 90.00 | 94.38 | 90.00 | ~ | 1.22-1.52 | ~ | -8.81 |
| IALG-$C_{30}$ | $P2/m$ | Our work | 4.30 | 7.47 | 9.92 | 90.00 | 99.43 | 90.00 | 1.90 | 1.22-1.52 | 189.64 | -8.81 |
| T-carbon | $Fd\bar{3}m$ | Our work | 7.52 | 7.52 | 7.52 | 90.00 | 90.00 | 90.00 | 1.50 | 1.42, 1.50 | 159.96 | -7.92 |
| | | GGA [26] | 7.52 | 7.52 | 7.52 | 90.00 | 90.00 | 90.00 | 1.50 | 1.42, 1.50 | 169.00 | -7.92 |
| carboneyane | $C2/m$ | Our work | 8.79 | 5.11 | 4.96 | 90.00 | 91.78 | 90.00 | 1.43 | 1.22-1.54 | 147.83 | -8.27 |
| | | GGA [28] | 8.82 | 5.09 | 4.96 | 90.00 | 91.49 | 90.00 | 1.43 | 1.22-1.54 | 140.00 | -8.27 |
| graphyne | $P6/mmm$ | Our work | 6.89 | 6.89 | ~ | 90.00 | 90.00 | 120.00 | ~ | 1.22-1.43 | ~ | -8.58 |
| | | GGA [30] | 6.90 | 6.90 | ~ | 90.00 | 90.00 | 120.00 | ~ | 1.22-1.43 | ~ | -8.58 |
| Diamond | $Fd\bar{3}m$ | Our work | 3.56 | 3.56 | 3.56 | 90.00 | 90.00 | 90.00 | 3.50 | 1.55 | 435.73 | -9.09 |
| | | GGA [31] | 3.57 | 3.57 | 3.57 | 90.00 | 90.00 | 90.00 | 3.50 | 1.55 | 418.00 | -9.09 |
| | | Exp. [32, 33] | 3.57 | 3.57 | 3.57 | 90.00 | 90.00 | 90.00 | 3.52 | 1.54 | 443.00 | ~ |
| Graphite | $P63/mmc$ | Our work | 2.46 | 2.46 | 6.80 | 90.00 | 90.00 | 120.00 | 2.24 | 1.42 | 266.66 | -9.21 |
| | | Exp. [33] | 2.46 | 2.46 | 6.70 | 90.00 | 90.00 | 120.00 | ~ | 1.42 | ~ | ~ |
| Graphene | $P6/mmm$ | Our work | 2.47 | 2.47 | ~ | 90.00 | 90.00 | 120.00 | ~ | 1.42 | ~ | -9.22 |

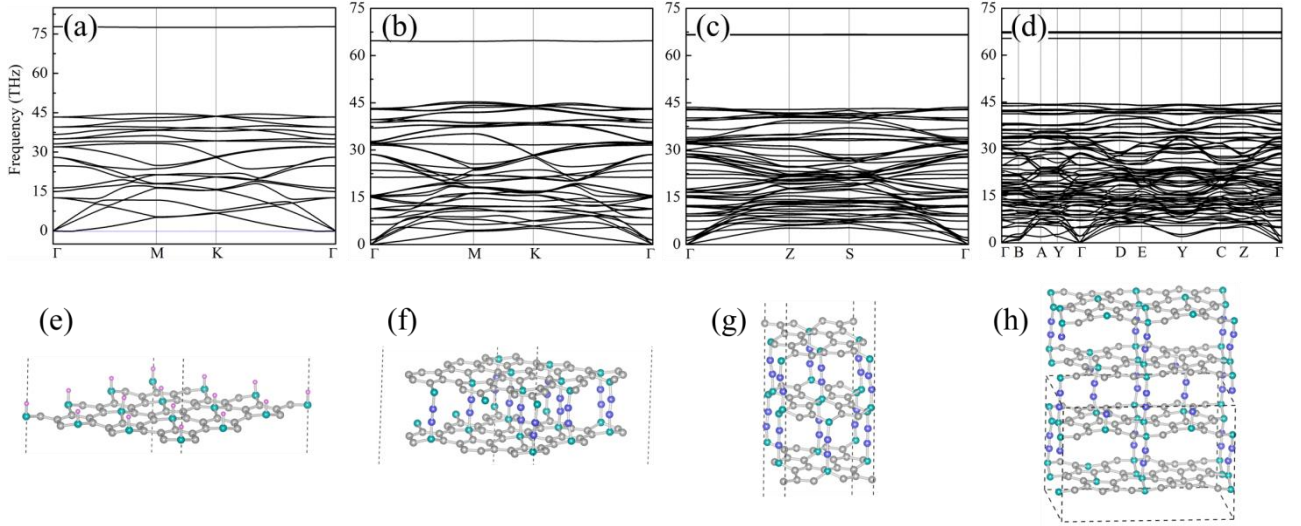

**Fig. 2.** (Color online) The dynamical and thermodynamic stabilities. Phonon spectra for (a) $C_6H$, (b) ALBG-$C_{14}$, (c) ALTG-$C_{22}$, and (d) IALG-$C_{30}$. Perspective views of the snapshots for those equilibrium structures of (e) $C_6H$, (f) ALBG-$C_{14}$, (g) ALTG-$C_{22}$, and (h) IALG-$C_{30}$ at the temperature of 300 K after 10 ps AIMD simulations.

After determining the structural stability and the magnetic ground state of these structures, we then study the electronic structures of these materials. Here, we also include $C_6H$ in the study. Since $C_6H$ [36] has a net magnetic moment of $1\,\mu_B$, it exhibits ferromagnetic band structure with the splitting spin-up and spin-down bands near the Fermi level (Fig. 3(a)). Furthermore, we find that the spin polarization is mainly localized on the three nearest-neighboring (NN) C atoms of the hydrogenated C atom, reflecting that the conjugated $\pi$-bonding network of graphene is destroyed by the adsorption of H atom. This leads to the electron localization and leaves unpaired electrons on the nonhydrogenated C atoms with the spin moment of 0.22 $\mu_B/C$ (see the inset of Fig. 3(a)). According to PBE calculations, the $C_6H$ holds an indirect band gap of 0.17 eV between the spin-up valence band maximum (VBM) at the K point and the spin-up conduction band minimum (CBM) at the $\Gamma$ point of the BZ. In comparison, the hybrid functional (HSE06) calculation gives a band gap of 0.87 eV (see Supplemental Material (SM) Fig. S1). From the PDOS in Fig. 3(b), one can see that the spin polarization mainly originates from the $p_z$-orbitals of the $sp^2$ C atoms, i.e., the three NN C atoms of the hydrogenated C atom, which is consistent with our results of charge density.

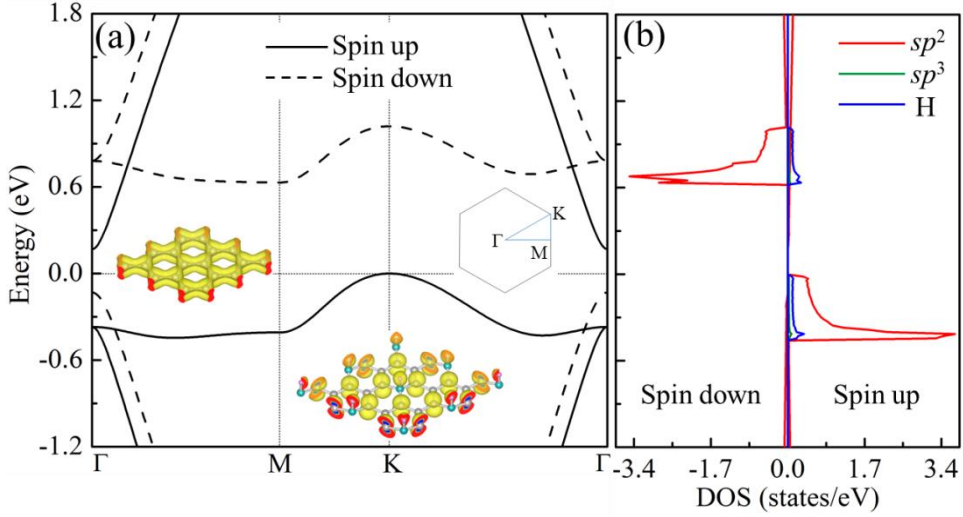

**Fig. 3.** (Color online) (a) The spin-polarized band structure and (b) projected density of states (PDOS) of $C_6H$ in the ground state (Néel). The charge densities of graphene and $C_6H$ at the VBM, and the Brillouin zone (BZ) are shown in the inset of panel (a).

For the three example carbon allotropes, the acetylenic linkages ($-C \equiv C-$) act as the hydrogen atom in $C_6H$ to connect the graphene layers. This will induce the magnetization in each layer, with a net magnetic moment of $1\,\mu_B$, similar to the case of $C_6H$. In ALBG-$C_{14}$, there are two cases respectively with interlayer FM and AFM couplings. We find that in the case of interlayer FM coupling, there will be a net magnetic moment of $2\mu_B$, which is due to the superimposed magnetic moments from the upper and lower layers. From Table 1, it can be seen that the energy with the interlayer AFM coupling is 3.5 meV/C lower than that with the interlayer FM coupling. So the magnetic ground state of ALBG-$C_{14}$ is the intralayer Néel state with the interlayer AFM coupling (without a net magnetic moment). Figure 4(a) shows the spin-polarized band structure for the ground state of ALBG-$C_{14}$, which is a semiconductor with a direct band gap of 0.32 eV (PBE calculations) between the VBM and the CBM at the $\Gamma$ point of the BZ. We can also observe that the band edges show flat-band characteristics, which is mainly due to the contribution of the spin-polarized $sp^2$ carbon atoms. Correspondingly, high peaks appear in the PDOS as shown in Fig. 4 (b). It is worth noting that the peaks of the $sp^2$ and $sp$ carbon atoms in PDOS have a large overlap, which implies that there is a strong interaction between them, so the $sp$ carbon atoms in ALBG-$C_{14}$ can help to mediate interlayer magnetic coupling. This can be verified from the charge density analysis (see the inset of Fig. 4(a)).

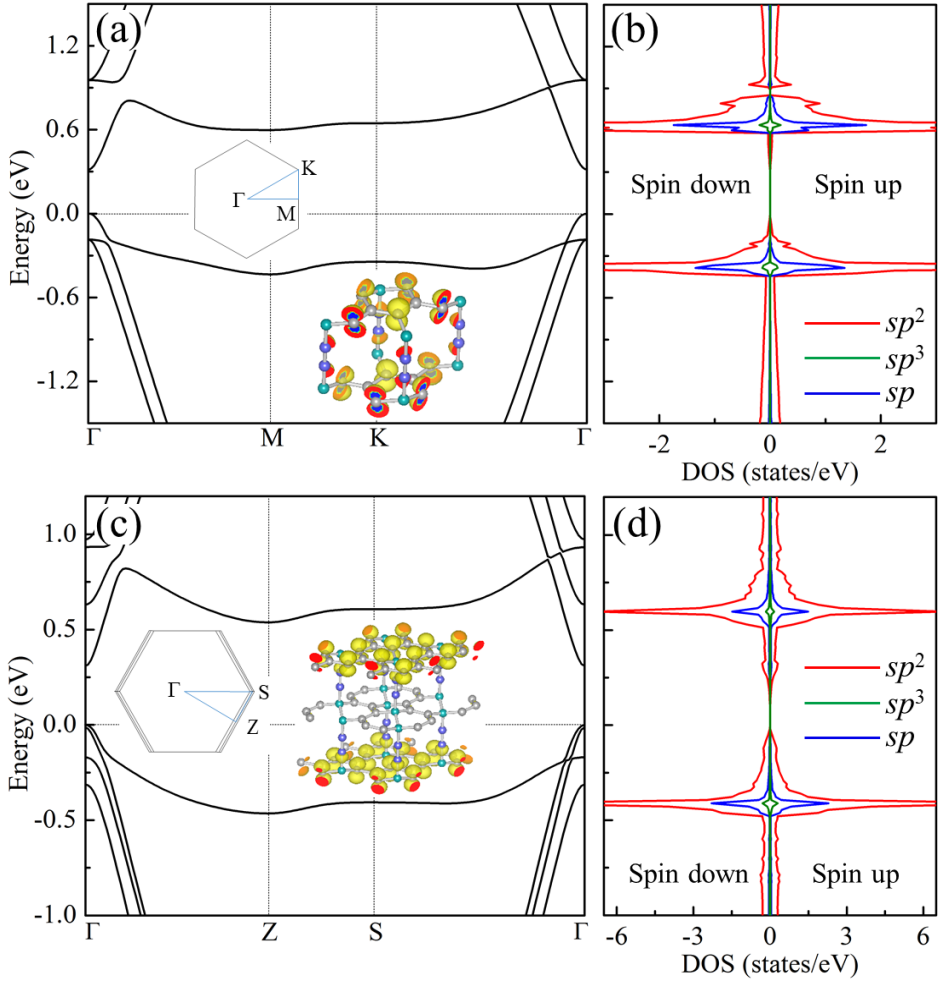

**Fig. 4.** (Color online) (a) The spin-polarized band structure and (b) PDOS of ALBG-$C_{14}$ in the ground state. (c) The spin-polarized band structure and (d) PDOS of ALTG-$C_{22}$ in the ground state. Their respective charge densities at the VBM, and BZ are shown in the inset of Fig. 4 (a) and 4 (c).

The triple-layer ALTG-$C_{22}$ can be viewed as obtained by connecting the A-B-A-stacked trilayer graphene with the acetylenic linkages. Compared with ALBG-$C_{14}$, ALTG-$C_{22}$ is special in that the carbon atoms on the A and B sublattices of the middle graphene layer are simultaneously connected by acetylenic linkages

($-C \equiv C-$) on both sides, such that $N_A^{p_z}$ is equal to $N_B^{p_z}$ in this layer. So there is no magnetic moment in the middle graphene layer, which can be confirmed by the charge density analysis at the VBM, as shown in the inset of Fig. 4(c). Similar to ALBG-C$_{14}$, ALTG-C$_{22}$ also has a net magnetic moment of $2\,\mu_B$ per cell by the superposition of top and bottom magnetic graphene layers, which is consistent with the mechanism that we propose. From Table 2, we find that the energy difference between the interlayer AFM state and the interlayer FM state of ALTG-C$_{22}$ is negligible. This is because the middle graphene layer is not spin polarized. This is equivalent to doubling the interlayer spacing (4.2-5.0 Å), resulting in a weak coupling between the two magnetic graphene layers (~10.0 Å). Therefore, the spin-polarized band structure (Fig. 4(c)) and the PDOS (Fig. 4(d)) of ALTG-C$_{22}$ are very similar to those of ALBG-C$_{14}$, but the energy gap of ALTG-C$_{22}$ becomes smaller (0.31 eV, PBE calculations).

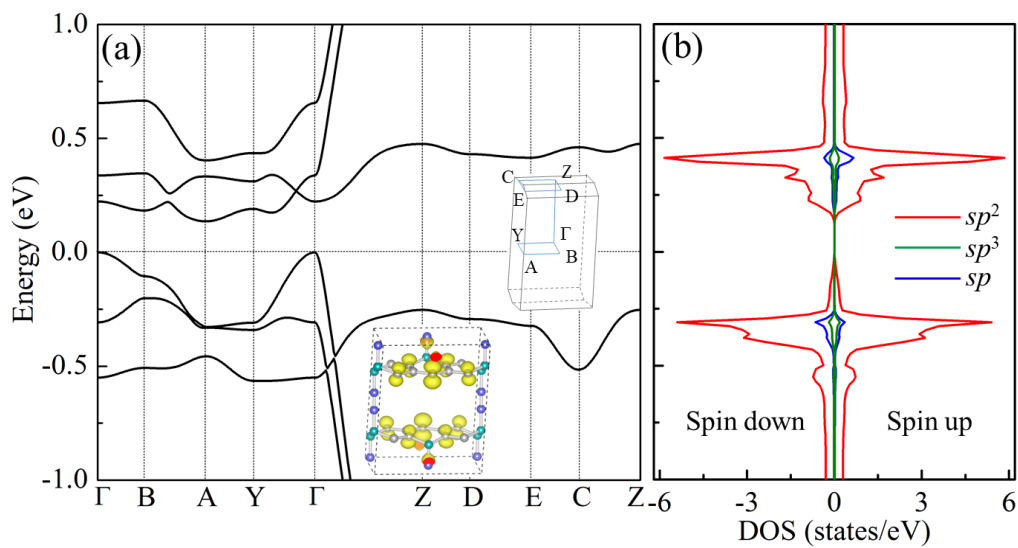

**Fig. 5.** (Color online) (a) The spin-polarized band structure and (b) the PDOS of IALG-C$_{30}$ in the ground state. The charge density of IALG-C$_{30}$ at the VBM, and the BZ are shown in the inset of panel (a).

Finally, we move to study the 3D bulk carbon material IALG-C$_{30}$, which can be derived from graphite by inserting appropriate acetylenic linkages. We note that there is a net magnetic moment of $2\,\mu_B$ in the interlayer FM state, which is obtained by superposing the net magnetic moments of the upper and lower graphene layers under the interlayer FM coupling. Since the $|N_A^{p_z} - N_B^{p_z}|$ numbers of the upper and lower graphene layers in the unit cell of IALG-C$_{30}$ (see Fig. 1(i)) both equal to $|5 - 4| = 1$, they each contribute 1 $\mu_B$ to the total magnetic moment. From Table 2, it is found that the energy of the state with interlayer AFM coupling is 1.6 meV/C lower than that with interlayer FM coupling, so IALG-C$_{30}$ has a ground state with the intralayer Néel and interlayer AFM couplings. Figure 5 shows the spin-polarized band structure and PDOS for the ground state of IALG-C$_{30}$. The system is a semiconductor with an indirect band gap of 0.14 eV (PBE calculations), and it has flat-band characteristics near the band edge. This is mainly due to the contribution of the spin-polarized *sp*$^2$ carbon atoms, consistent with the charge density analysis shown in the inset of Fig. 5(a).

Recently, many experiments have been performed to study the room-temperature (RT) ferromagnetic behaviors in various forms of carbon-based materials [37-40]. The discovery of RT carbon magnets has stimulated the search for more magnetic carbon materials with FM [41-44] and AFM [45-47] couplings. Nearly 30 years ago, an FM crystal structure (ferrocarbon) containing equal numbers of *sp*$^2$ and *sp*$^3$ hybridization was predicted by Ovchinnikov *et al*. [48]. However, Pisani *et al.* found that this structure is unstable and undergoes a barrier-free transition to a non-magnetic *sp*$^3$ hybridized carbon network by using hybrid functional calculations [49]. This implies that the carbon structure formed by linking the same sublattice of graphene between the buckled graphene sheets is unstable. So far, there are very few reports of the intrinsic magnetism in all-carbon crystal structures, although many heuristic design schemes of carbon allotropes and experimentally known carbon structures have been reported [50-53]. Our previous work [25] found that the acetylenic linkages $(-C \equiv C-)$ can connect the carbon atoms on the same sublattices of graphene to form a stable carbon network structure, which provides a good opportunity to induce the intrinsic magnetism in the graphene-based all-carbon materials. Furthermore, the conjugated $\pi$-bonding networks are widely distributed in real materials [54, 55], which will allow our theoretical proposal to be applied to more systems. In order to facilitate the experimental preparation, we provide a possible scheme to synthesize the magnetic carbon structures we proposed. With the development of the current functionalized graphene technology [56-58], it may be possible to stack multilayer brominated graphene sheets [59], insert acetylene molecules between the stacked layers, and finally deacidify to form various 2D or 3D magnetic carbon materials, which are not only limited to ALBG-$C_{14}$, ALBG-$C_{22}$, and IALG-$C_{30}$.

## 4. Conclusions

In summary, we propose a unique magnetization mechanism similar to Ovchinnikov's rule to guide the design of a series of novel magnetic all-carbon crystal structures. The idea is to combine different stacked graphene layers via acetylenic linkages ($-C \equiv C-$) to construct a class of 2D and 3D carbon network structures. We confirm the effectiveness of this mechanism by studying several concrete examples. We show that ALBG-$C_{14}$ and ALTG-$C_{22}$ are 2D antiferromagnetic semiconductors (respective direct bandgaps of 0.32 and 0.31 eV) with intralayer Néel state and interlayer AFM coupling. And IALG-$C_{30}$ is a 3D bulk magnetic carbon allotrope, with properties similar to those of ALBG-$C_{14}$ and ALTG-$C_{22}$. Our work offers a promising design scheme for magnetic carbon materials, and the mechanism may also apply to other systems with π-bonding networks beyond carbon.

## Declaration of competing interest

The authors declare that they have no known competing financial interests or personal relationships that could have appeared to influence the work reported in this paper.

## Acknowledgements

This work was supported by the National Key R&D Program of China (Grants No. 2019YFA0308603 and No. 2017YFA0302903), the National Natural Science Foundation of China (Grants No. 11774422, No. 11774424, and No. 11804039), the Beijing Natural Science Foundation (Grant No. Z200005), the CAS Interdisciplinary

Innovation Team, the Fundamental Research Funds for the Central Universities, and the Research Funds of Renmin University of China (Grants No. 16XNLQ01 and No. 19XNLG13), and the Singapore Ministry of Education AcRF Tier 2 (MOE2017-T2-2-108). Y.G. was supported by the Outstanding Innovative Talents Cultivation Funded Programs 2021 of Renmin University of China. Computational resources were provided by the Physical Laboratory of High Performance Computing at Renmin University of China.